\documentclass[preprintnumbers,amsmath,amssymb]{revtex4}
\usepackage{amsmath,amsthm}
\usepackage{amssymb,latexsym}
\usepackage{graphicx}
\usepackage{setspace}
\usepackage[mathscr]{eucal}
\begin{document}
\title{Aerodynamic Stability of Satellites in Elliptic Low Earth Orbits}
\author{Matthew Bailey, Brian Ladson\\Stefan C. Mancas, Bogdan Udrea, Uchenna Umeadi}

\affiliation{Embry-Riddle Aeronautical University,\\ Daytona Beach, FL. 32114-3900, U.S.A.}

\begin{abstract}
Topical observations of the thermosphere at altitudes below $200 \, km$ are of great
benefit in advancing the understanding of the global distribution of mass,
composition, and dynamical responses to geomagnetic forcing, and momentum 
transfer via waves. The perceived risks associated with such low altitude and
short duration orbits has prohibited the launch of Discovery-class missions.
Miniaturization of instruments such as mass spectrometers and advances in the
nano-satellite technology, associated with relatively low cost of nano-satellite
manufacturing and operation, open an avenue for performing low altitude
missions.
%
The time dependent coefficients of a second order non-homogeneous ODE which describes the motion have a double periodic shape.
Hence, they will be approximated using Jacobi elliptic functions. Through a change
of variables the original ODE will be converted into Hill's ODE for stability analysis using Floquet theory.

We are interested in how changes in the coefficients of the ODE affect the
stability of the solution. The expected result will be an allowable range of
parameters for which the motion is dynamically stable. A possible extension of
the application is a computational tool for the rapid evaluation of the
stability of entry or re-entry vehicles in rarefied flow regimes and of
satellites flying in relatively low orbits.
\end{abstract}
\maketitle
\section{Introduction}

Non-dimensional aerodynamic coefficients are a measure of performance of a
vehicle moving through a fluid at different speeds. The non-dimensional
aerodynamic coefficients salient for this proposal are the pitch damping
$(C_{m_{\dot{\alpha}}} + C_{m_q})$ and pitch stiffness $C_{m_{\alpha}}$
coefficients. The
pitch coefficients determine the pitch stability of the vehicle.

In general, the pitch damping and stiffness coefficients are functions of the 
geometry of the vehicle, angle of attack, the rate of the angle of attack, 
and the pitch rate. For a vehicle operating in flow regimes with a Knudsen 
number larger than
unity particle-surface interactions have a large influence on the pitch
coefficients. These interactions are defined by energy accommodation 
coefficients.
Examples of vehicles operating in this regime are relatively low flying
spacecrafts and re-entry vehicles.

In the preliminary stages of a low flying or re-entry mission study, a large 
level of uncertainty exists in the values used for the non-dimensional aerodynamic 
coefficients. The reasons for the high level uncertainty are 
multiple. The most important reason is the fact that little it is known about
the energy accommodation coefficients. The second most important uncertainty is
the composition of the atmosphere throughout the time interval of interest.

\begin{figure}
\begin{center}
\includegraphics[scale=0.32]{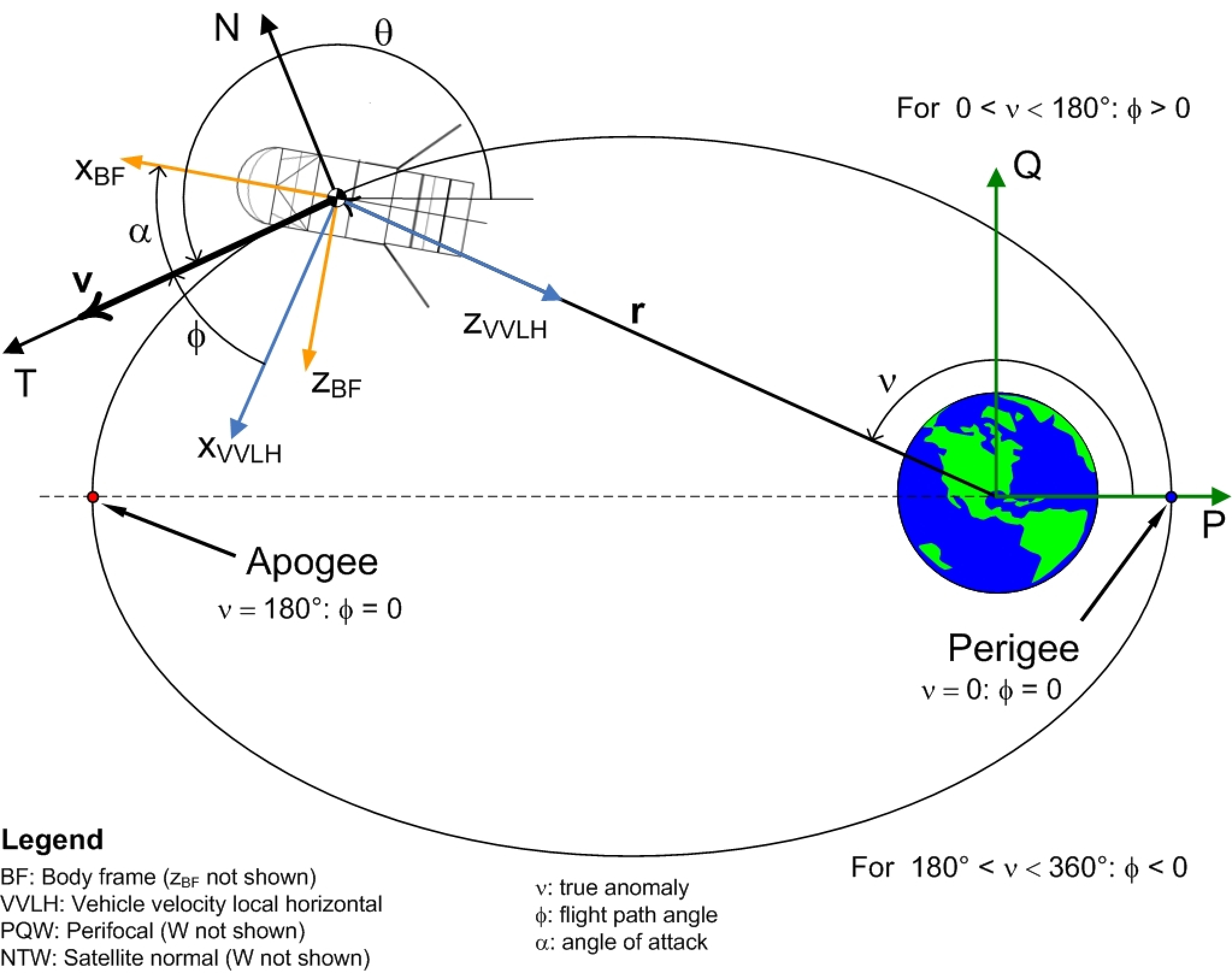}
\caption{\em Coordinate systems employed in the derivation of the equation of
         describing the pitching motion of the spacecraft.}
\label{fig:coord_systems_orbit}
\end{center}
\end{figure}

The purpose of this paper is to analyze the pitch stability of a
vehicle flying in the rarefied flow regime and determine ranges of pitch
damping and pitch stiffness coefficients for which the pitch motion is stable. For this purpose we use analytical methods based on  Floquet theory \cite{Chicone}, and numerical methods developed in MATLAB \cite{Chapra} to solve the following second order linear ODE 
\begin{equation} \label{ODE}
\ddot{\alpha}(t)-\big(M_{q}(t)+M_{\dot{\alpha}}(t)\big)\dot{\alpha}(t)-M_{\alpha}(t) \alpha(t)=\ddot\theta(t),
\end{equation}
where $\alpha(t)$ is the angle of attack, $M$ is the pitching moment, 
and $\theta$ is the equivalent of the pitch angle and they are all functions of time.
The following notation has been used: $M_q=\frac{1}{J}\frac{\partial M
}{\partial q}$, $M_{\dot{\alpha}}=\frac{1}{J}\frac{\partial M }{\partial
\dot{\alpha}}$, $M_{\alpha}=\frac{1}{J}\frac{\partial M }{\partial \alpha}$,
$J$ being the moment of inertia. The relationship between the $M$ coefficients
of the ODE and the non-dimensional pitching moment coefficients is given in
Eq. \eqref{pitch_coeffs}.
The coordinate systems employed in the definition of the angles are presented
in Fig.~\ref{fig:coord_systems_orbit}.

Due to their observed double periodic shape, the periodic coefficients $M_{q}(t)+M_{\dot{\alpha}}(t)$ (the damping term), and $M_{\alpha}(t)$ (the stiffness), will be approximated using Jacobi elliptic functions $sn{(\zeta,m)}, cn{(\zeta,m)}/ dn{(\zeta,m)}$, where independent variable $\zeta$ is a function of time, and $m$ is the modulus of the Jacobi functions used \cite{Bowman}. Of interest is how changes in the damping and stiffness coefficients will affect the stability of the pitch motion. The expected result will be an allowable range of parameters for which the motion is dynamically stable. One possible application of the technique is a tool for the rapid study of the stability of entry or re-entry vehicles. 

\section{Procedure}

\subsection{Orbit}

This section refers specifically to the analysis of CubeSat mission. One of the mission's requirements is that the Dipping Thermospheric Explorer 
(DiPTE) 
CubeSat shall operate in an elliptical orbit of $700 \,km$ apogee and $200\,km$ 
perigee.
The other mission requirement relevant to this proposal is that the 
pointing knowledge of the
payload sensitive axis in the ram direction shall be within 
$0.1^\circ$ ($3 \sigma$)
full cone and the pointing control shall be $2.0^\circ$ ($3 \sigma$) full cone
in the ram direction.

The orbit requirement is derived from the need to perform measurements of
density and temperature and of wind direction and magnitude of relevance
for the propagation of various types of waves in the thermosphere. The attitude
knowledge and control requirement is derived from the accuracy constraints of
the payload.

\subsection{Approximation of Aerodynamic Moments}

It has been noticed numerically that the time dependent coefficients of the second order non-homogeneous ODE which describes the motion of the satellite have a double periodic shape.

The forcing function $\ddot\theta(t)$ of Eq. \eqref{ODE} is obtained from the equations for the components of the velocity vector expressed in the satellite body frame and the satellite normal frame, and is due to the rotating frame in which the angle of attack was defined. This function is known and is also periodic and obtained from taking twice the time derivative of
\begin{equation} \label{theta}
\theta(\nu)=\arctan\Big(-\frac{\cos \nu+e \cos 2\nu}{\sin \nu+ e \sin 2\nu}\Big),
\end{equation}
where $\nu$ is the true anomaly, and $e$ is the eccentricity of the orbit given by $e=\frac{R_{apo}-R_{peri}}{R_{apo}+R_{peri}}$. Here, the distance from focus to apoapsis $R_{apo}=Rad_{Earth}+700$, and the the distance from focus to periapsis $R_{peri}=Rad_{Earth}+200$, with the radius of the Earth being the equatorial Earth radius $R_{Earth}=6378.1363 \, Km$.

In order to accomplish the overall goal of analyzing the pitch stability of this space vehicle flying in the rarefied flow regime and ultimately determine the ranges of pitch damping and pitch stiffness coefficients for which the pitch motion is stable, we used a preliminary method of an improved approximation of the density data using a least squares approximation of Jacobian elliptic functions \cite{Chapra, Bowman}.

Let $y_i$ = $y(t_i)$ be the set of ($n$ density $\rho(t)$, or any other periodic function) data points that we are trying to model using a general least squares model. Then
\begin{equation} \label{summ}
y=\sum_{j=0}^m {a_j}{z_j}+\epsilon,
\end{equation}
where $z_j$ = $sn${($\zeta_j,\kappa_j$)} are $m+1$ base elliptic functions, $\zeta_j$ = $j\omega_0t$ are the harmonics, $\kappa_j$ the modulus of the elliptic sine function used, and $\epsilon$ is the error that we want to minimize.

Note that instead of $sn$, one could use $cn, dn$ or any other combination (that works). It is important to have analytical expressions for the moments $M_q(t)$, $M_{\dot\alpha}(t)$, and $M_\alpha(t)$ since these coefficients determine the pitch stability of the vehicle, and they are needed to compute the angle of attack $\alpha(t)$ by solving Eq. \eqref{ODE}.

Using this method and having analytical expressions for $\rho(t)$ in terms of elliptic functions, the damping and stiffness moments are obtained via
\begin{align}
M_{q}(t)&=\frac{1}{2 J v(t)}Q(t)S_{ref}l^2_{ref}C_{m_q}\notag\\
M_{\dot{\alpha}}(t)&=\frac{1}{2 J v(t)}Q(t)S_{ref}l^2_{ref}C_{m_{\dot{\alpha}}}\notag\\
M_{\alpha}(t)&=\frac{1}{J}Q(t)S_{ref}l_{ref}C_{m_\alpha},
\label{pitch_coeffs}
\end{align}
where $Q(t) =  \frac{\rho(t)v(t)^2}{2}$ is the dynamic pressure, $\rho(t)$ is the mass density of the air, $v(t)$ is the speed, $l_{ref}$ is the reference length (0.1 $m$), and $S_{ref}$ is the reference surface area (0.01 $m^2$), and $C_{m_q}$ , $C_{m_{\dot{\alpha}}}$, $C_{m_{\alpha}}$ are the pitch damping and stiffness coefficients ( with $C_{m_q}+C_{m_{\dot{\alpha}}}=-10$, and $C_{m_{\alpha}}=-1.635$ respectively).

Using this notation, Eq. \eqref{summ} can be written in matrix form as
\begin{align} \label{Y_matrix}
Y=ZA+E, 
\end{align}
where the bases matrix $Z$ is given by
\begin{align}
Z=\begin{bmatrix}
 1 & sn{({\omega_0}{t_1}, \kappa_1)} & ... &  sn{(m{\omega_0}{t_1}, \kappa_m)} \\ 
 1 & sn{({\omega_0}{t_2}, \kappa_1)} & ... &  sn{(m{\omega_0}{t_2}, \kappa_m)} \\
 . & . &  & . \\
 . & . &  & . \\
 . & . &  & . \\
 1 & sn{({\omega_0}{t_n}, \kappa_1)} & ... &  sn{(m{\omega_0}{t_n}, \kappa_m)}
\end{bmatrix}
\label{Z}
\end{align}
Therefore, the sum of squares of residuals is formulated as follows
\begin{equation} \label{error}
S_r=\sum_{i=1}^n \Big(y_i-\sum_{j=0}^m {a_j}{z_{ji}}\Big)^2.
\end{equation}
By taking $m + 1$ partial derivatives with respect to $a_j$, the normal equation is $A = (Z^tZ)^{-1}Z^tY$, where $A = (a_0\;a_1\;...\;a_m)^t$ are the unknown coefficients of the base functions that we are trying to find, and $y = (y_1\;y_2\;...\;y_n)^t$ are raw the density data points.

In order to accomplish a least squares model of the periodic density function, first we determine the period of the data which is $T$ = 5615.2 sec. This value was determined from the original Pitch Dynamics MATLAB code. Using this value for period, $\omega_0$ is found from $\omega_0=2\pi/T$. 

Let the Jacobian elliptic integral \cite{Bowman} be defined as 
\begin{align}
u=\int_0^x \! {\frac{dt}{ \sqrt{1-t^2}\sqrt{1-{k^2}{t^2}}}},
\end{align}
then $x = sn(u, k)$. When $x = 1$, the complete elliptic integral becomes
\begin{equation} \label{K}
K=\int_0^1 \! {\frac{dt}{ \sqrt{1-t^2}\sqrt{1-{k^2}{t^2}}}},
\end{equation}
which has the quarter period $K = T/4$. Because the above integral that defines the full elliptic sine function cannot be solved analytically, MATLAB was used to systematically find the modulus that matched with the raw data's period of $T$ = 5615.2 sec. The modulus found using Eq. \eqref{K} was $k = 0.4405$. In this procedure we assumed that all base functions have the same modulus $\kappa = k^2$.

The data used for approximating the density as an elliptic sine approximation comes from a mass density profile text file developed by Rick Doe, SRI International \cite {SRI}. The MATLAB code reads in the altitude and density columns for the file in order to conduct proper calculations using the appropriate data. However, before calculations begin, the altitudes of interest are extracted with their corresponding density values.  Of the original 486 data points from the text file, only 251 points are useful.  These 251 points acquire half the period, which means there are a total of 502 data points to fill one orbit. The true anomaly values range from 0 to 359 degrees, but we wanted to evaluate our motion starting from the apogee. Therefore, $\pi$ must be added to each true anomaly value (after being converted into radians) in our range. Based on our data points, a step size is created accordingly to fit the data.  This step size is determined by taking the largest value of the range (359 degrees) and dividing it by one less the total number of data points (501). The altitude at the perigee and apogee is 200 km and 700 km, respectively. The radius of the Earth at these altitudes can be determined, which can lead to values for the semi-major axis, eccentricity, and orbit period.  The orbital position and speed provide calculations using Kepler's formulas, and this will be explained later. 

In order to begin the elliptic sine approximation, the data points were sorted accordingly based on the fact that motion is starting from the apogee. The $Z$ matrix is developed via Eq. \eqref{Z}.  As discussed above, the $A$ matrix is $(Z^tZ)^{-1}Z^tY$.  Eq. \eqref{Y_matrix} develops the matrix for the density approximation using elliptic sine functions. The sum of the squares residual that was calculated using the elliptic functions method was $S_r = 7.9210 10^{-13}$ which is a far better method of approximation of the data than the use of the cubic spline approximation method that had an error $S_r = 2.6839 10^{-8}$.

\begin{figure}
\begin{center}
\includegraphics[scale=0.35]{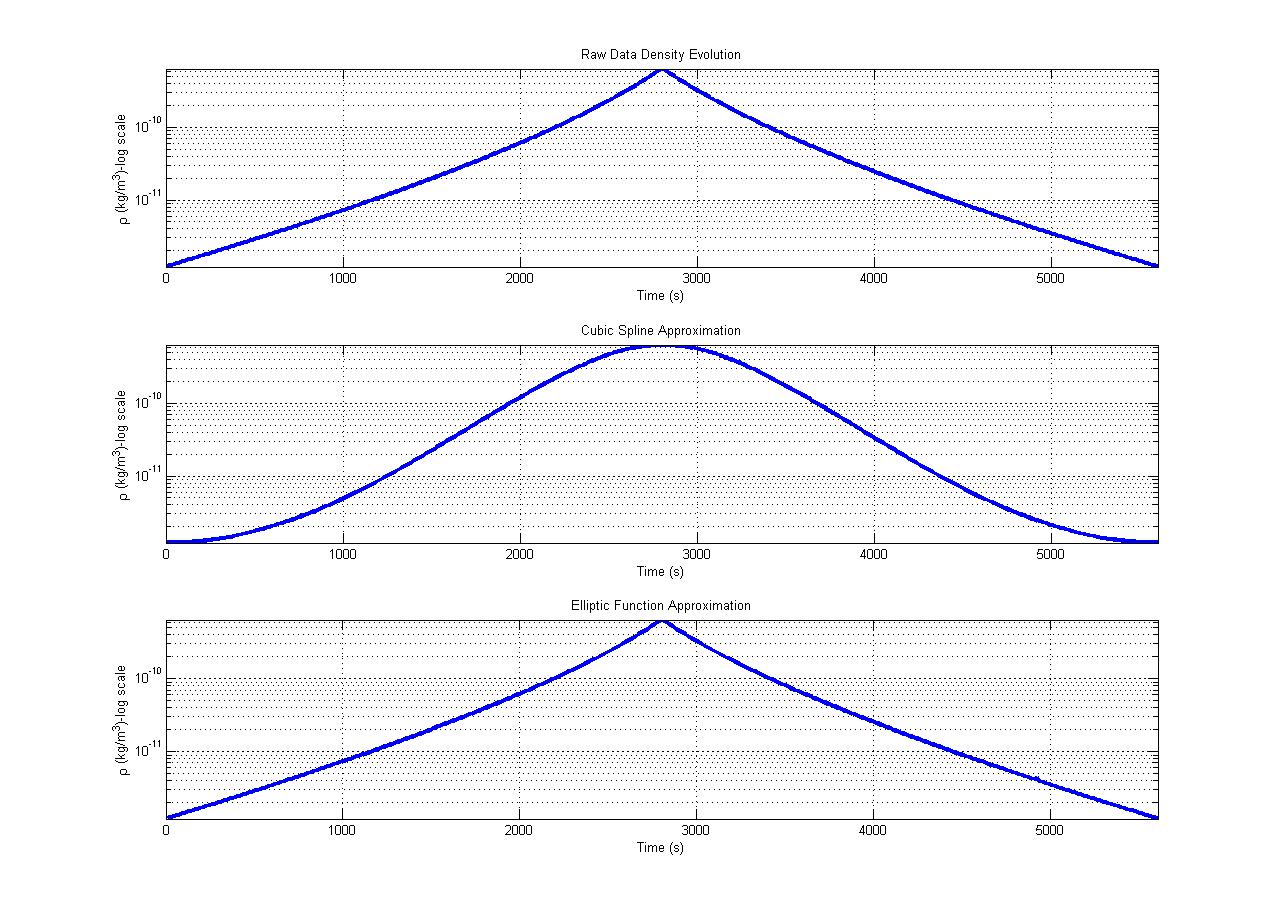}
\caption{\em Density raw data (top), approximation using splines (middle), approximation using elliptic functions (bottom).}
\label{fig:density}
\end{center}
\end{figure}

By consulting the comparison in Fig.~\ref{fig:density}, it is notable that the percent difference between the errors calculated for the two methods was 199.988\%. Regardless that the percent difference error is so large, the errors of these approximations are relatively small in comparison to the raw data on the logarithmic scale.

\begin{figure}
\begin{center}
\includegraphics[scale=0.25]{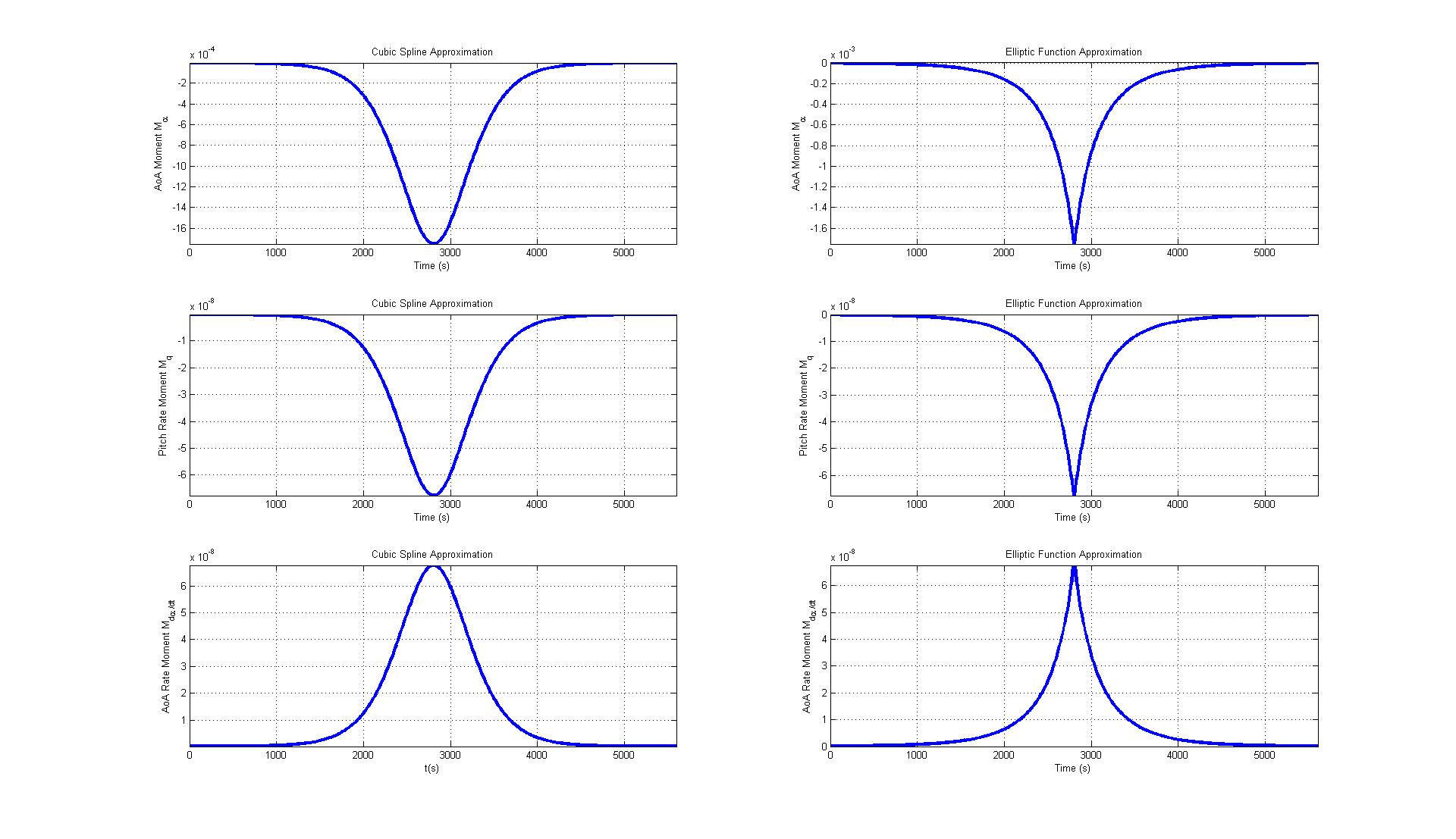}
\caption{\em Aerodynamic moments, approximation using splines (left), approximation using elliptic functions (right).}
\label{fig:moments}
\end{center}
\end{figure}

Since we have analytical expression for the density, by using Eq. \eqref{pitch_coeffs}, we obtained the moments $M_q(t)$, $M_{\dot\alpha}(t)$, and $M_\alpha(t)$. The comparison between the cubic spline approximation and the elliptic sine approximation of the aerodynamic moments are shown in Fig.~\ref{fig:moments}.

\subsection{Pitch Dynamics}

The $700 \, km \times200 \,km$ elliptic orbit makes the pitch dynamics an interesting 
dynamic problem due to the fact that the density varies by three order of
magnitudes between the apogee and the perigee. The density at the apogee is so
low that the dynamics of the pitch motion is very close to that of a double
integrator. At the perigee the density is sufficient to provide enough pitch
stiffness and some pitch damping.

\begin{figure}
\begin{center}
\includegraphics[scale=0.5]{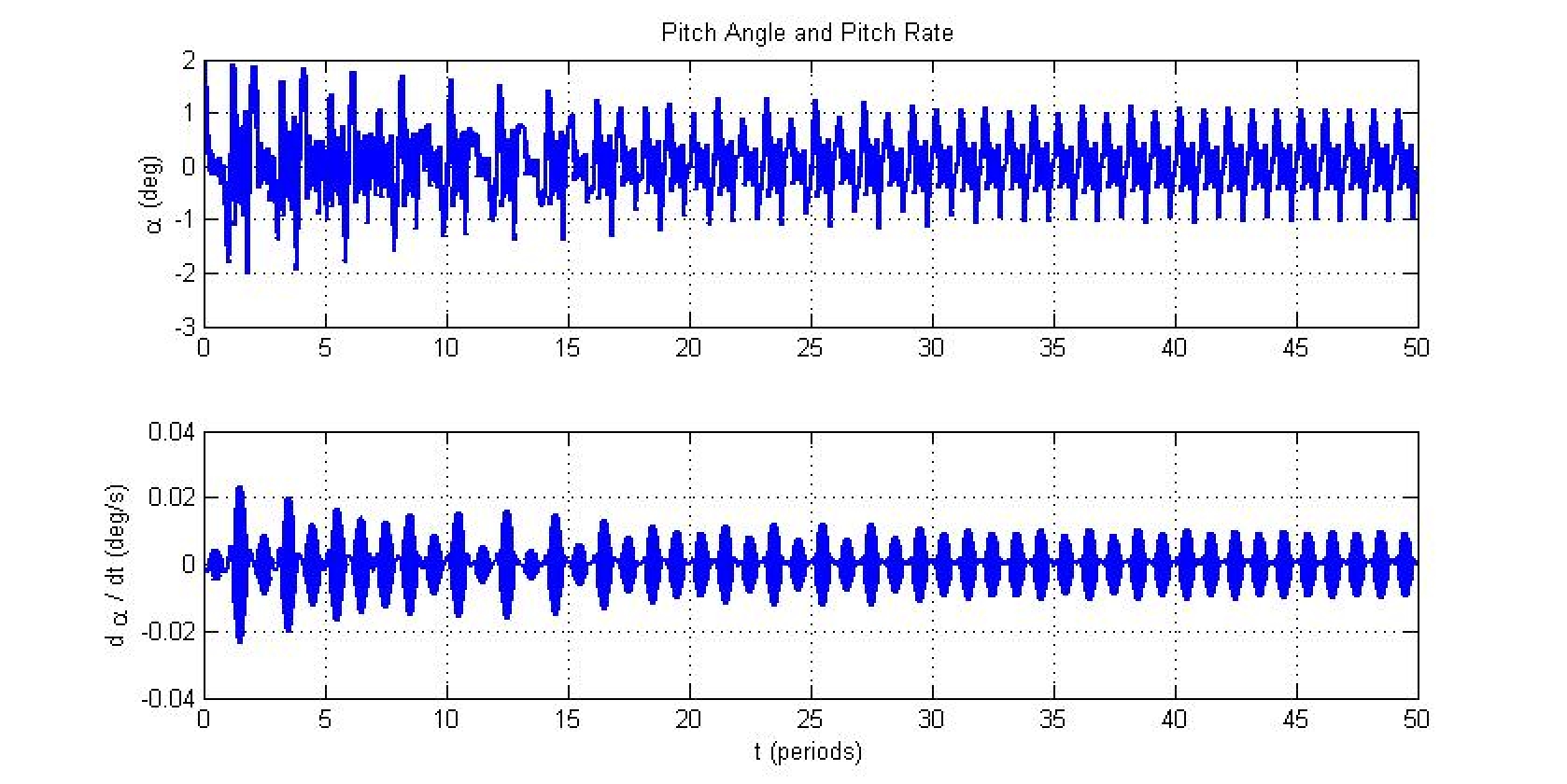}
\includegraphics[scale=0.5]{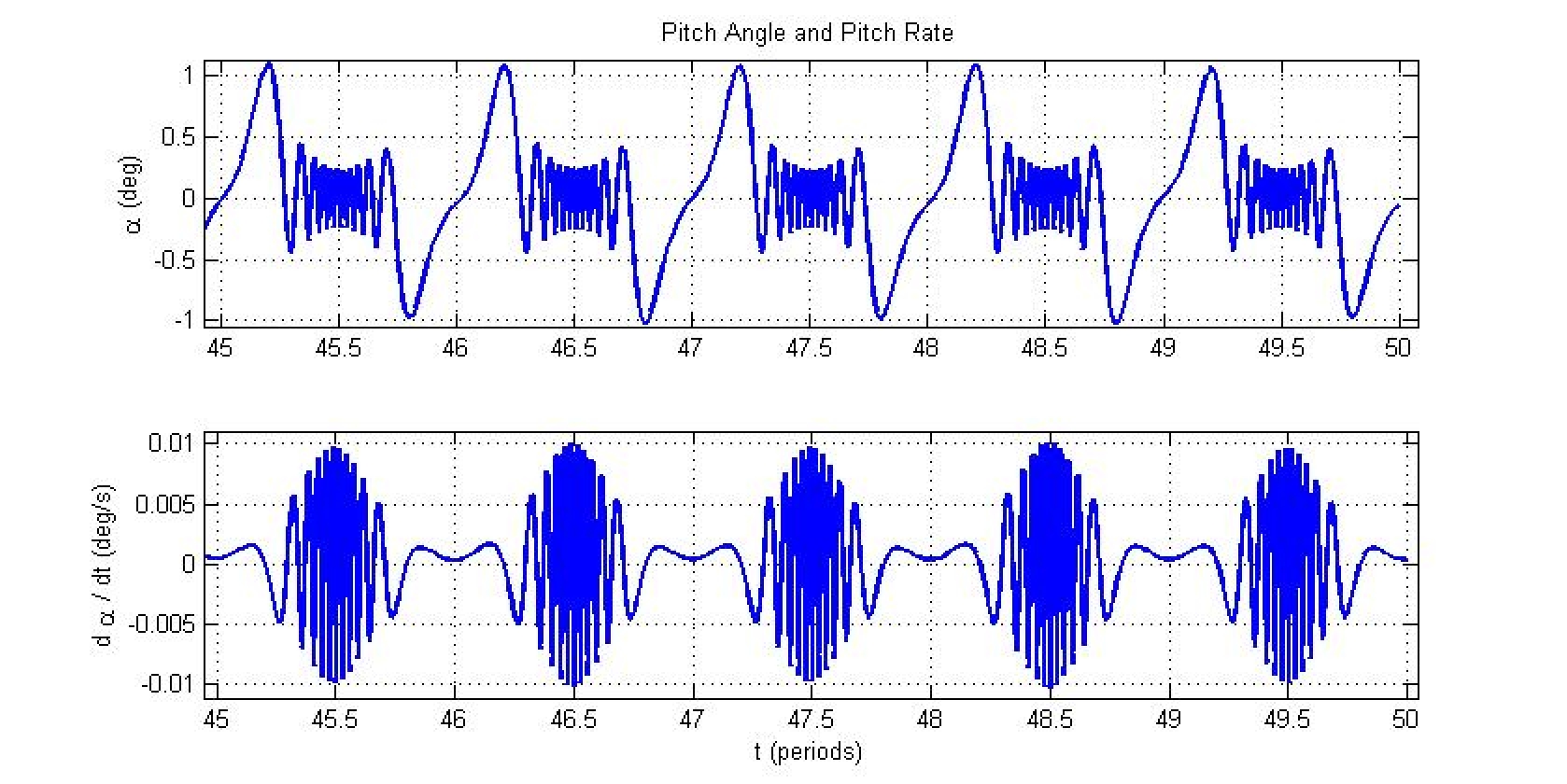}
\caption{\em Time history of the pitch angle and pitch rate for 50 orbits (top) and a
         zoom in for the last five orbits (bottom).}
\label{fig:pitch_history}
\end{center}
\end{figure}

\begin{figure}
\begin{center}
\includegraphics[scale=0.52]{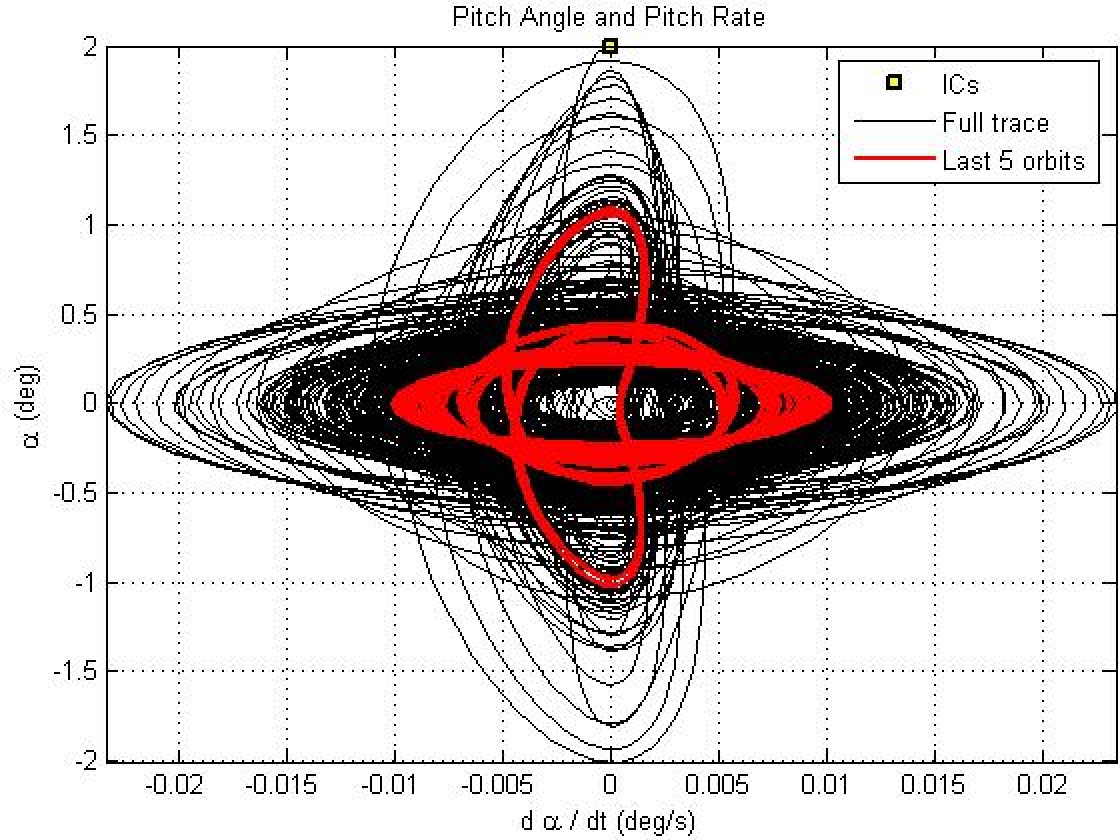}
\caption{\em Phase plot diagram of the pitch dynamics.}
\label{fig:pitch_history_phase}
\end{center}
\end{figure}

Numerical integration of the homogeneous ODE \eqref{ODE} using MATLAB, describing the pitch motion shows the expected
behavior. The results of the integration are presented in
Fig.~\ref{fig:pitch_history}. The moments used initially are based on the cubic spline approximation. The initial conditions used are pitch angle
of $\alpha_0 = 2^{\circ}$ and pitch rate of $\dot{\alpha}_0 = 0$. The pitch 
settles in an oscillatory motion with oscillations at the perigee with an
amplitude of $0.35^{\circ}$ well within the $2{^\circ}$ full cone requirement.
It is interesting to take a look at the pitch motion in the phase plane, i.e.,
pitch angle vs. pitch rate, shown in Fig.~\ref{fig:pitch_history_phase}. 
The phase plane diagram is similar to that of a
limit cycle motion. Since the motion of the pitch angle and pitch rate indicates a quasiperiodic motion, in the phase plane the motion is limited to an attractor. 

\subsection{Pitch Stability}
The purpose of this paper is to analyze the pitch stability of the CubeSat flying in the rarefied flow regime and to determine its ranges of pitch damping and pitch stiffness coefficients for which the pitch motion is stable based on elliptic functions. In order to do so, we performed an analysis using analytical methods delineated in Floquet theory \cite{Chicone}, together with numerical methods developed in MATLAB. We will now discuss the methods and algorithms utilized in the construction of operational MATLAB code in order to solve the nonhomogeneous (ODE) as denoted in Eq. \eqref{ODE}.

First, after many different tries, we have established that the most appropriate ODE solver script for our problem is provided was the \texttt{ode113} solver. This is primarily used for solving non-stiff differential equations by the means of variable order method. This solver integrates the system of differential equations over the time period given by $t_{span}$, and evaluates the solution starting with the initial conditions given by $y_0$. The output is the  column vector $y_{ODE}$ with corresponding time vector $t_{ODE}$.

To find the right-hand side of Eq. \eqref {ODE}, we differentiated twice Eq. \eqref{theta} and we obtained

\begin{equation}
\dot\theta(\nu,\dot\nu)=\arctan\Big(-\frac{1+2e^2+3e\cos \nu}{1+e^2+2e\cos \nu}\dot\nu\Big),
\end{equation}

\begin{equation}\label{ddots}
\ddot\theta(\nu,\dot\nu,\ddot\nu)=\frac{1+2e^2+3e\cos \nu}{1+e^2+2e\cos \nu}\ddot\nu+\frac{e(e^2-1)\sin \nu}{(1+e+2e\cos \nu)^2}\dot\nu^2.
\end{equation}

It is important to note that the true anomaly $\nu(t)$, along with its first two time derivatives, are also functions of time. Their expressions will also be approximated using elliptic sine functions as follows. Since via Eq. \eqref{summ}
\begin{equation} \label{niu}
\nu(t)=a_0+\sum_{j=1}^m {a_j}sn(j\omega_0t),
\end{equation} then by differentiating, 

\begin{equation} \label{niudot}
\dot{\nu}=\omega_0\sum_{j=1}^m {ja_j\sqrt{[1-sn^2(j \omega_0 t)][1-k^2 sn^2(j\omega_0t)]}},
\end{equation}
and

\begin{equation} \label{niuddot}
\ddot{\nu}=\omega_0^2\sum_{j=1}^m {j^2a_jsn(j\omega_0t)[2k^2sn^2(j\omega_0t)-(1+k^2)]}.
\end{equation}
The values of  $\cos \nu$, and $\sin \nu $ necessary in obtaining $\ddot\theta$ were calculated using the Keplerian equations that govern elliptically orbiting bodies. From the equation of radius $R$
\begin{equation}\label{ar}
R=\frac{a(1-e^2)}{1+e\cos \nu},
\end{equation} where $a=\frac{R_{apo}+R_{peri}}{2}$ is the semimajor axis, we find $\cos \nu$ and subsequently $\sin \nu$, and we substitute them in Eq. \eqref{ddots}. 251 data points data points delineating the evolution the satellite altitude from its apogee to its perigee were concatenated in order to represent the altitude evolution for a full orbit and then were populated into a vector that served as a raw data for our elliptic sine approximation. Furthermore,  the speed $v(t)$ was determined via Eq. \eqref{sp}.
\begin{equation}\label{sp}
v(t)=\sqrt{\frac{\mu}{R} \Big(\frac{1+e^2}{1+e\cos \nu(t)}\Big)},
\end{equation} 
where $\mu=3.986004415\,10^5 \,\frac{Km^3}{s^2}$ represents the Earth's gravitation parameter.

Using the analytical expression for $\nu(t)$ in terms of the elliptic sine function, Eq. \eqref{theta}, and the MATLAB \texttt{ode113} solver, we solved numerically both the homogenous and nonhomogenous ODE Eq. \eqref{ODE}. The results are graphically plotted and analyzed below.
\begin{figure}
\begin{center}
\includegraphics[scale=0.25]{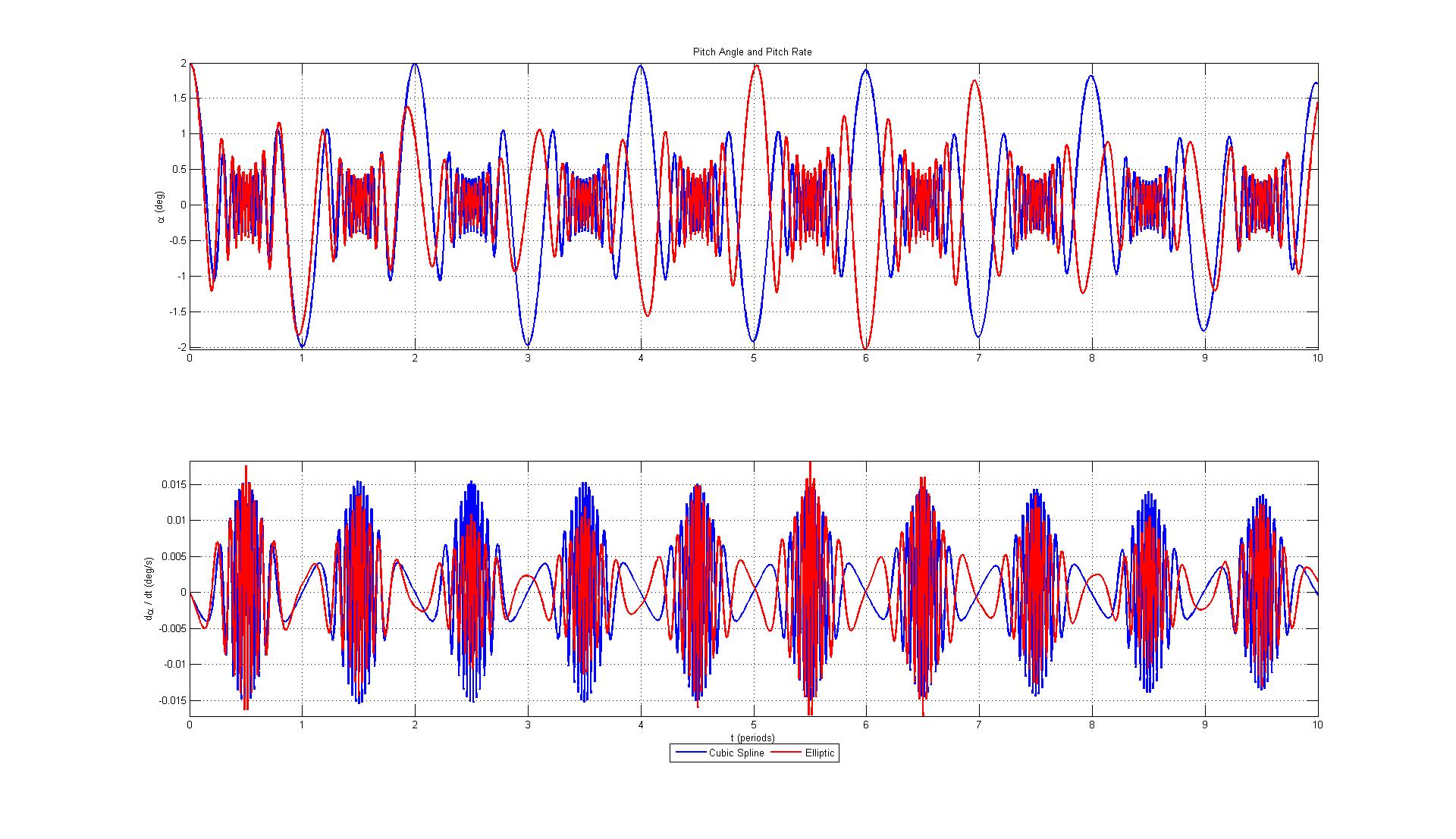}
\caption{\em Pitch angle and pitch rate overlay for the homogeneous ODE.}
\label{fig:pitch_homogeneous}
\end{center}
\end{figure}
Fig.~\ref{fig:pitch_homogeneous} shows the homogeneous solutions of $\alpha(t)$ and $\dot\alpha(t)$ over a full 10 periods (orbits). The figure actually shows an overlay of the solution as determined using cubic spline approximations and elliptic sine approximations. It is important to note that because these solutions are homogeneous, (the right hand side of the ordinary differential equation is null) they do not take into account the true anomaly approximations which solely appear in the forcing function $\ddot\theta$. It is evident, as previously noted, that the newly found results from the elliptic sine approximations closely simulate the cubic spline approximation. This is a testament to the accuracy of the results and the reliability of the approximation method as employed. Fig.~\ref{fig:pitch_homogeneous} also testifies to the accuracy of the results and the consistency of the findings.

\begin{figure}
\begin{center}
\includegraphics[scale=0.25]{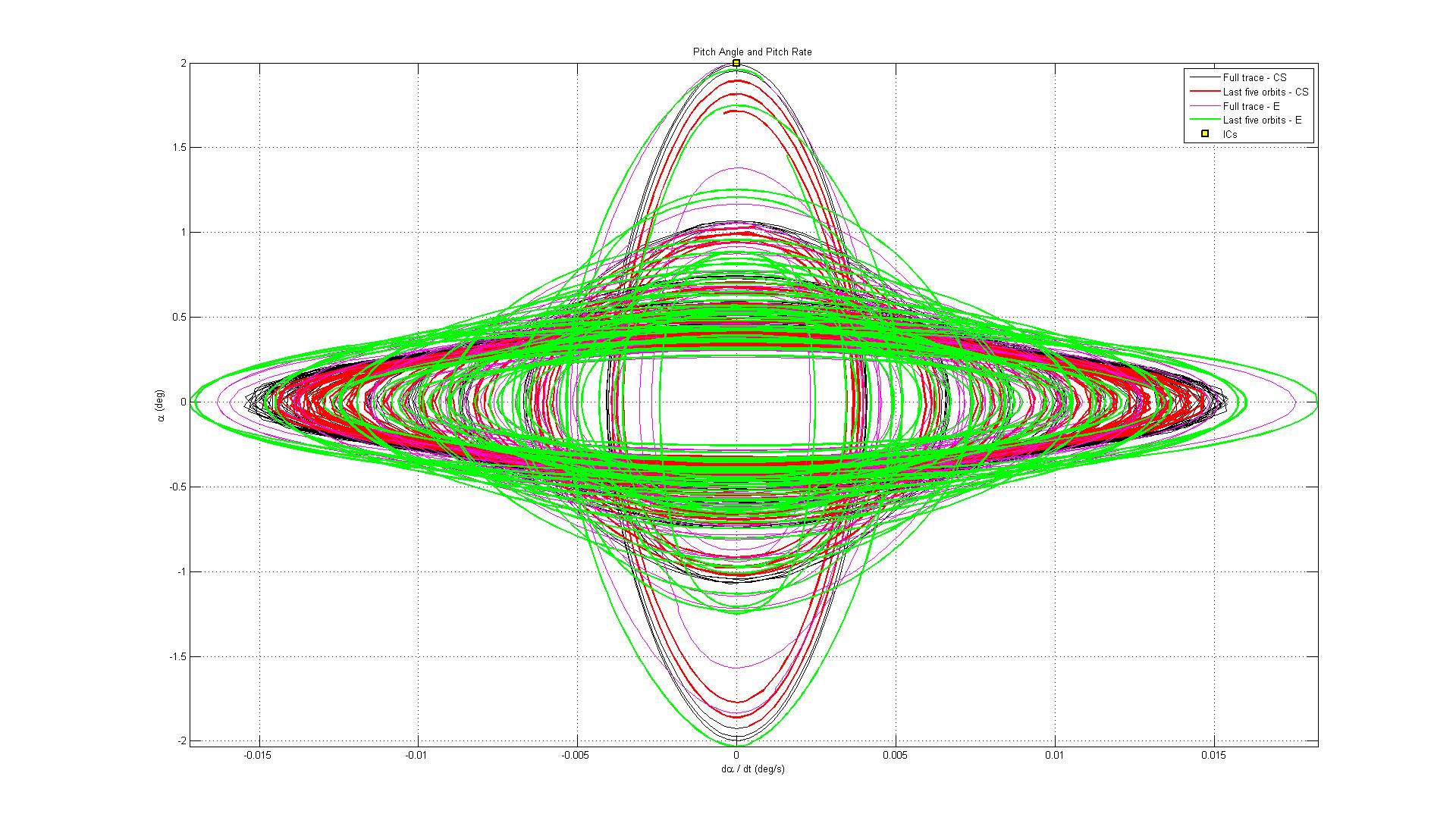}
\caption{\em Phase plot overlay of the 2 methods solving the homogeneous ODE.}
\label{fig:phaseplot_homogeneous}
\end{center}
\end{figure}

Fig.~\ref{fig:phaseplot_homogeneous} shows the phase plot overlay of the same information (the homogeneous solution from the cubic spline and elliptic sine approximations).
The behavior of both approximated solutions were similar. The shape and predicted envelope of the pitch angle and pitch rate are comparable. The solutions produced by the two methods of interest begin to diverge; however when the non-homogeneous solutions are explored, the subsequent figures show that these solutions and the variations are more evident.
\begin{figure}
\begin{center}
\includegraphics[scale=0.25]{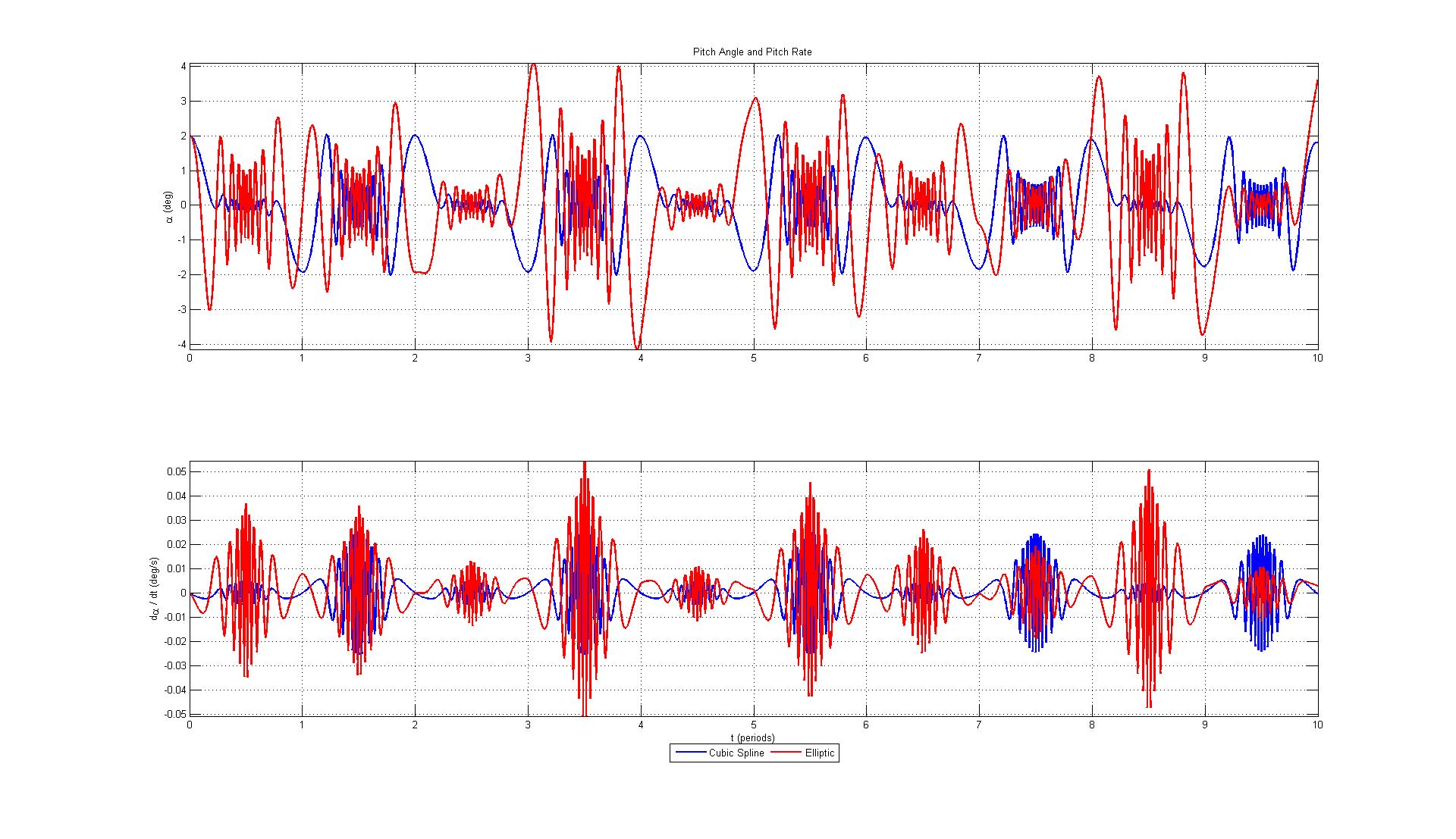}
\caption{\em  Pitch angle and pitch rate overlay for the nonhomogeneous ODE.}
\label{fig:pitch_nonhomogeneous}
\end{center}
\end{figure}
At first glance, it is readily noticeable that both solutions exhibit the same general characteristics, namely a double periodic behavior with beats, as seen in Fig.~\ref{fig:pitch_nonhomogeneous}. The simulation of longitudinal dynamics shows a damped oscillatory behavior with these beats that line up without a phase shift that would otherwise suggest inaccuracy in the results. Although the amplitude of the elliptic sine results is consistent with the previous findings with respect to the comparable cubic spline results in that they are larger. These larger amplitudes can be attributed to the more direct approximation that is done through the elliptic sine function that yields results that are more accurate yet less controlled.

Looking now at the phase plot overlay of the same results, we can develop a similar conclusion.
\begin{figure}
\begin{center}
\includegraphics[scale=0.25]{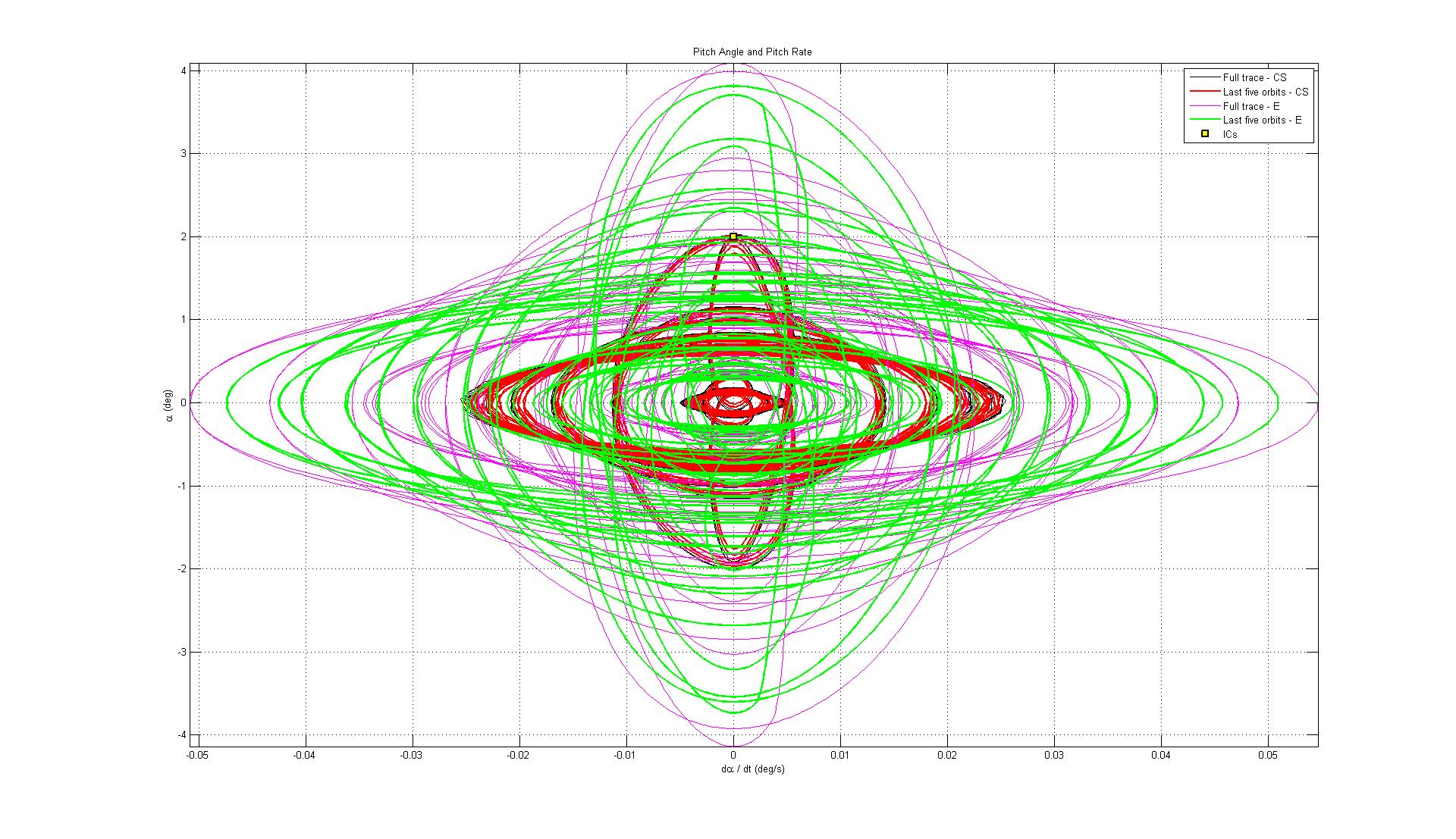}
\caption{\em Phase plot overlay of the 2 methods solving the nonhomogeneous ODE.}
\label{fig:phaseplot_nonhomogeneous}
\end{center}
\end{figure}
Though the general attitude (behavioral shape) of the competing solutions is similar, there is a distinct issue of difference when it comes to the magnitude of the results as displayed in Fig.~\ref{fig:phaseplot_nonhomogeneous}. The elliptic sine results yield values that are on a scale of almost a factor of 2 both for the pitch angle and the pitch rate. This indicate an error in the calculations, such as derivations for $\dot\nu$ and $\ddot\nu$, or may just be the reality of what the elliptic sine approximations yield. The final conclusion is yet to be made as the researchers continue to investigate the results and possible alterations to the method used to solve the ODE.

The results indicate a strong correlation between the cubic spline and elliptic sine approximations, yet do suggest that there may be discrepancies in the method used for finding and approximating the true anomaly values that are employed in the resolution of the ODE via elliptic sine methods.

Further investigation will be considered on the use of other Jacobian periodic elliptic functions in order explore the aerodynamic longitudinal pitch stability of the satellite. Floquet theory is the next step of the comprehensive process in developing a full analytical and numerical evaluation of the stability regimes of the CubeSat.

\subsection{Dynamical Systems Analysis Using Floquet Theory}

The parameters
associated with the dynamics of the motion of the of a CubeSat class mission
flying in a $700 \, km \times 200\, km$ orbit have been computed with a direct simulation Monte Carlo code or extrapolated from existing data. In this project we also determined the stability regions of Eq. \eqref{ODE} describing the one degree of freedom attitude dynamics in low altitude elliptic orbits using Floquet therory.

Once the damping and stiffness coefficients are known, by the change of variables
\begin{equation}\label{trans}
\alpha(t)=\beta(t) e^{-\frac{1}{2}\int_0^t (M_{q}(s)+M_{\dot{\alpha}}(s))ds},
\end{equation}
the homogeneous Eq. \eqref{ODE} will be converted into Hill's equation, Eq.\eqref{Hill}. This equation was introduced by George W. Hill in his studies of the motion of the Moon. Roughly speaking the motion of the Moon can be viewed as a harmonic oscillator in a periodic gravitational field.
Since the analytic solutions of Eq. \eqref{ODE} are not known, to analyse the stability, we will be using Floquet theory. Here, $b(t)$ is also a periodic function such that $b(t)=b(t+\hat{T})$ for some $\hat{T}$ which needs to be found; and it will act as an energy source for the system. 
\begin{equation}\label{Hill}
\ddot{\beta}(t)+b(t)\beta(t)=0
\end{equation}
In order to solve Eq. \eqref{Hill}, $b(t)$ must be determined first. Using Eq. \eqref{ODE}, set $a_1=-[M_{q}(t)+M_{\dot{\alpha}}(t)]$ and $a_2=-M_{\alpha}(t)$ in order to apply the transformation given by \eqref{trans}. The ODE now possesses the form $\ddot{\alpha}(t)+a_1\dot{\alpha}(t)+a_2\alpha(t)=\ddot{\theta}(t)$. After developing the appropriate derivatives of Eq. \eqref{trans}, substituting into the transformed ODE, and completing some algebraic manipulation, the transformed ODE yields the expression provided by \eqref{Hill} such that
\begin{equation}\label{b}
b(t)=-M_{\alpha}(t)+\frac{1}{2}\frac{d}{dt}[M_{q}(t)+M_{\dot\alpha}(t)]-\frac{1}{4}[M_{q}(t)+M_{\dot\alpha}(t)]^2
\end{equation}

$\ddot{\theta}(t)$ will transform to $\gamma(t)$, however Hill's ODE solves for a homogenous differential equation. Therefore $\gamma(t)$ does not need to be determined in this case. To continue with proving the stability of the satellite, $b(t)$ must be calculated. According to \eqref{pitch_coeffs},
\begin{align}
M_{q}(t)+M_{\dot{\alpha}}(t)=\frac{S_{ref}{l_{ref}}^2}{4J}(C_{m_q}+C_{m_{\dot{\alpha}}})v(t)\rho(t) \notag\\
\frac{d}{dt}[M_{q}(t)+M_{\dot\alpha}(t)]=\frac{S_{ref}{l_{ref}}^2}{4J}(C_{m_q}+C_{m_{\dot{\alpha}}})\frac{d}{dt}[v(t)\rho(t)]
\label{bcalc}
\end{align}
where $\frac{d}{dt}[v(t)\rho(t)]=\dot{v}(t)\rho(t)+v(t)\dot{\rho}(t)$. Using \eqref{bcalc} and the expression for $M_{\alpha}(t)$ from \eqref{pitch_coeffs}, and letting an arbitrary constant $A=\frac{S_{ref}{l_{ref}}^2}{4J}$, then
\begin{align}\label{bbfinal}
b(t)=-2A \frac{C_{m_{\alpha}}}{l_{ref}}\rho(t)[v(t)]^2+
\frac{A}{2}(C_{m_q}+C_{m_{\dot{\alpha}}})[\dot{v}(t)\rho(t)+v(t)\dot{\rho}(t)]-\frac{A^2}{4}(C_{m_q}+C_{m_{\dot{\alpha}}})^2[v(t)]^2[\rho(t)]^2
\end{align}
Furthermore, $\dot{\rho}(t)$ is obtained via \eqref{summ} and by using the identities, $\frac{d}{du}sn(u)=cn(u) dn(u)$, and $cn(u)=\sqrt{1-sn^2(u)}$, $dn(u)=\sqrt{1-k^2 sn^2(u)}$, i.e., see also Eq. \eqref{niudot}, with $\nu$ replaced by $\rho$. 

Using linear stability of dynamical systems, we will transform Eq. \eqref{Hill} into a first order equivalent system
\begin{eqnarray}
\left[\begin{array}{cc}
\dot{x_1}\\
\dot{x_2}\\
\end{array}\right]=\left[\begin{array}{cc}
0& 1\\
-b(t) & 0\\
\end{array}\right]\left[\begin{array}{cc}
x_1\\
x_2\\
\end{array}\right].
\end{eqnarray}
Applying linear systems theory, the stability of the zero solution of the linear periodic system can be analyzed.  From the scalar second order linear differential equations, if $\ddot{u}+p(t)\dot{u}+q(t)u=0$, and the Wronskian of the two solutions $u_1$ and $u_2$ being defined as
 \begin{eqnarray}
 W(t):= \det \left[\begin{array}{cc}
u_1(t) & u_2(t)\\
\dot{u_1}(t) & \dot{u_2}(t)\\
\end{array}\right],
 \end{eqnarray}
 then by Liouville's Lemma,
 \begin{equation} 
W(t)=W(0) e^{-\int_0^t p(s)ds}
\end{equation}
Let the characteristic multipliers for Hill's Equation be denoted by $\rho_1$ and $\rho_2$ and note they are roots of the characteristic equation
\begin{align}
\rho^2 - (\operatorname{tr}\Phi(T)) \rho + \det \Phi(T) = 0.
\end{align}
For notaional convenience let us set $2\phi=\operatorname{tr}\Phi(T)$, to obtain the equivalent characteristic equation
 \begin{equation}\label{10}
 \rho^2-2 \phi \rho+1=0
 \end{equation}
 whose solutions are given by 
  \begin{equation}\label{99}
 \rho_{1,2}=\phi \pm \sqrt{\phi^2-1}.
 \end{equation}
Even though the solutions $u$ are not known explicitly, we have that the Floquet multipliers $\rho_{1,2}$ satisfy Eq. \eqref{10}
where $\phi=\frac 12 (\dot{u_1}(T)+\dot{u_2}(T))$.
The characteristic Floquet exponents $\mu_{1,2}$ are given by $\rho_{1,2}=e^{\mu_{1,2}T}$, and consequently using Eq. \eqref{10} 
\begin{align}
\mu_1+\mu_2&=0\notag\\
\cosh \mu_1 T&=\phi
\end{align}
Although $\phi$ is not known explicitly, it is useful to characterize the properties of $\rho_{1,2}$, or $\mu_{1,2}$ in terms of $\phi$.

Hence, by Eq. \eqref {99} we have the following cases:
\begin{itemize}
\item[(i)] if $\phi>1$, then $\rho_{1,2}$ are distinct positive real numbers such that $\rho_1 \rho_2=1$. Thus, we may assume that $0<\rho_1<1<\rho_2$, with $\rho_1=1/\rho_2$, and there us a real number $\mu>0$ (a characteristic exponent) that $e^{\mu \hat{T}}=\rho_2$ and $e^{-\mu \hat{T}}=\rho_1$. Then, there is a fundamental solutions set of the form $e^{-\mu t}p_1(t)$, and $e^{\mu t}p_2(t)$ where the real functions $p_{1,2}(t)$ are $\hat{T}$ periodic. In this case the zero solution is unstable;
\item[(ii)] if $\phi<-1$, then $\rho_{1,2}$ are both real and both negative. Also, since $\rho_1 \rho_2=1$ then we may assume that $\rho_1<-1<\rho_2<0$, with $\rho_1=1/\rho_2$.  Thus, there us a real number $\mu>0$ (a characteristic exponent) such that $e^{2\mu \hat{T}}=\rho_1^2$ and $e^{-2\mu \hat{T}}=\rho_2^2$. As in the case (i), there is a fundamental solutions set of the form $e^{\mu t}q_1(t)$, and $e^{-\mu t}q_2(t)$ where the real functions $q_{1,2}(t)$ are $2\hat{T}$ periodic. Again, the zero solution is unstable;
\item[(iii)] if $-1<\phi<1$, then $\rho_{1,2}$ are complex conjugates  with nonzero imaginary parts. Since $\rho_1\bar{\rho_2}=1$, we have $|\rho_1|=1$, and therefore both characteristic multipliers lie on the unit circle in the complex plane. Because $\rho_{1,2}$ have nonzero imaginary parts, one of this characteristic multipliers, say $\rho_1$, lies in the upper half plane. Thus, there is a real number $\theta$ with $0<\theta \hat{T}<\pi$ and $e^{i\theta \hat{T}}=\rho_1$. In fact, there is a solution of the form $e^{i \theta t}\big(r(t)+i s(t)\big)$, where $r(t),s(t)$ are both $\hat{T}$ periodic functions. Hence, there is a fundamental solutions set of the form $r(t) \cos (\theta t)- s(t) \sin(\theta t)$, $r(t) \sin (\theta t)+ s(t) \cos(\theta t)$. In particular, the zero solution is stable but not asymptotically stable. Also, the solutions are periodic if and only if there are relatively prime integers $m$ and $n$ such that $2\pi m=n \theta \hat{T}$. If such integers exist all solutions have period $n\hat{T}$. If not, then the solutions are quasi-periodic.
\end{itemize}
Certain curves of the form $\phi=\pm 1$ separate parameters regimes where unbounded solutions exist, i.e., $|\phi|>1$, from regions  where all solutions are bounded , i.e., $|\phi|<1$. We have just proved the following facts for Hill's equation, and the results are summarized in the following \underline{Lemma}, \cite{Chicone}, which we will act as the stability/instability criterion in determining the parameters' regimes for which the solution is stable/unstable.

\underline{Lyapunov Lemma}: \textit{If $\int_0^{\hat{T}} b(t) dt \leq \frac {4}{ \hat{T}}$, then all solutions of the Hill's equation \eqref{Hill} are bounded. In particular the trivial solution is stable.}\\

\begin{figure}
\begin{center}
\includegraphics[scale=0.45]{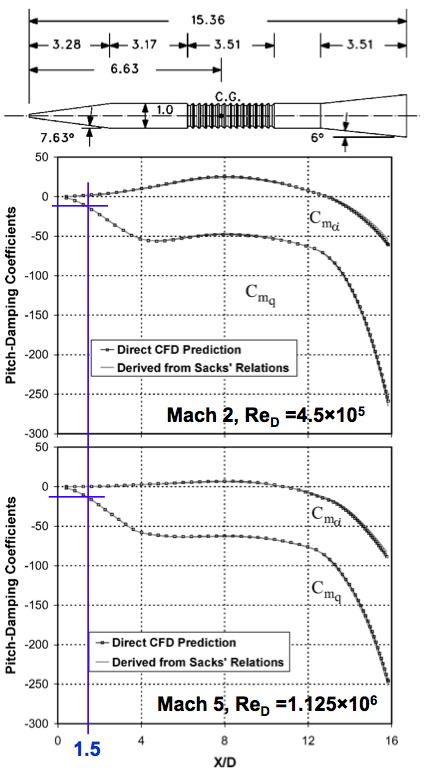}
\caption{\em Pitch-damping coefficients based on $x/D$ for Mach 2 and 5.}
\label{fig:damping}
\end{center}
\end{figure}

\begin{figure}
\begin{center}
\includegraphics[scale=0.25]{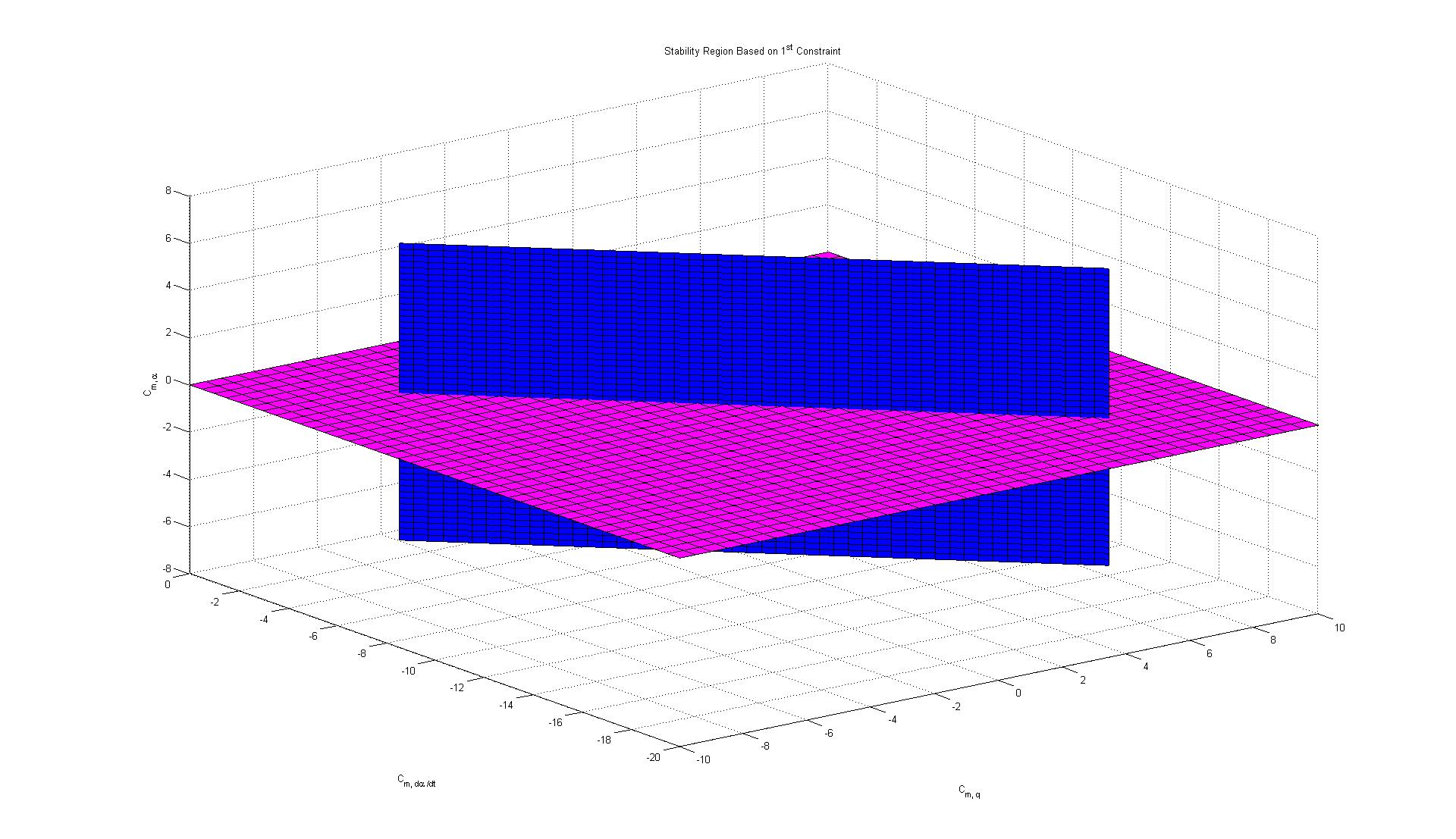}
\caption{\em Stability region based on the first constraint, $C_{m_{q}}+C_{m_{\dot{\alpha}}}=-10$.}
\label{fig:condition1}
\end{center}
\end{figure}

\begin{figure}
\begin{center}
\includegraphics[scale=0.25]{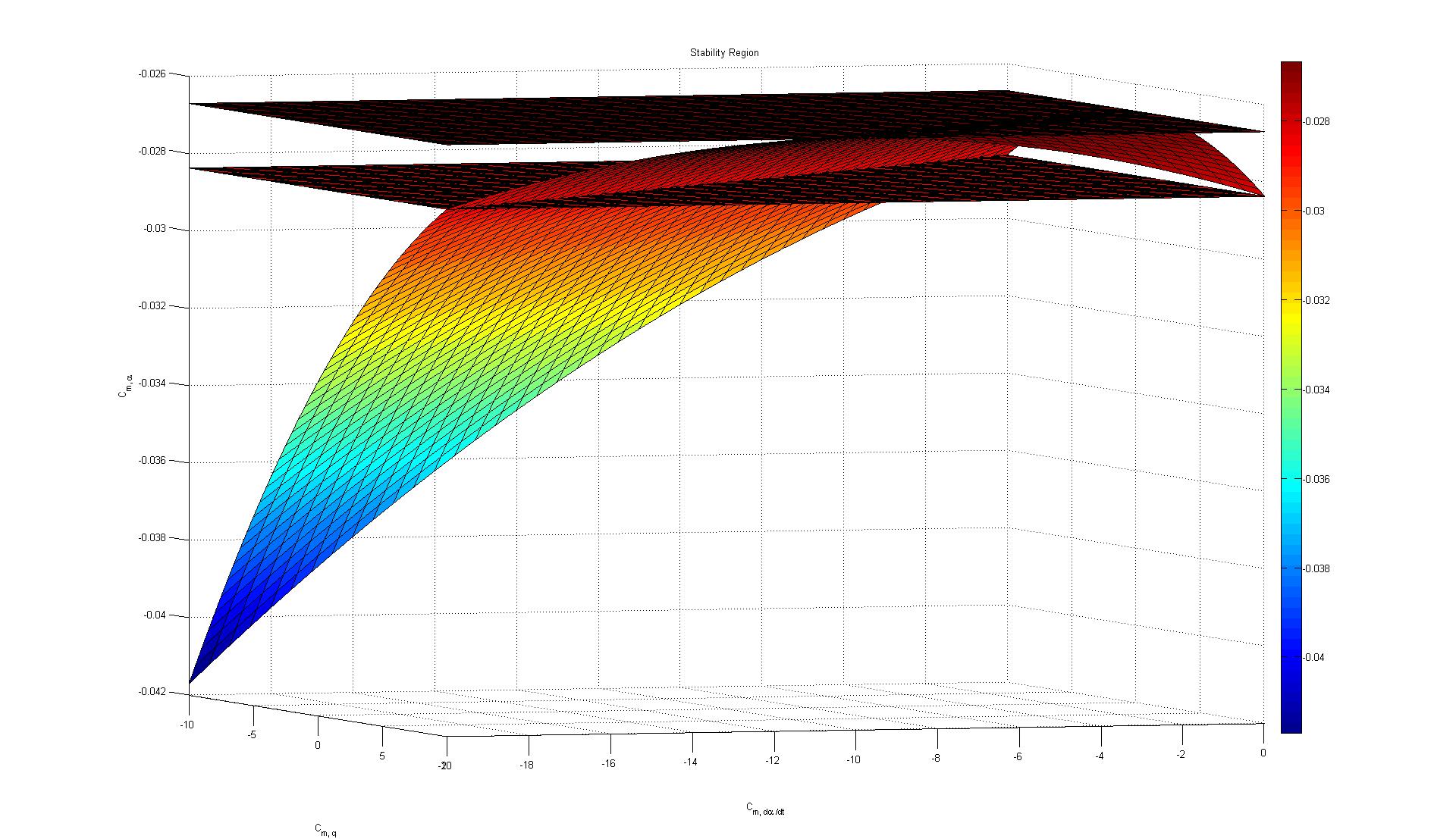}
\caption{\em Stability region surface.}
\label{fig:stability}
\end{center}
\end{figure}

\begin{figure}
\begin{center}
\includegraphics[scale=0.25]{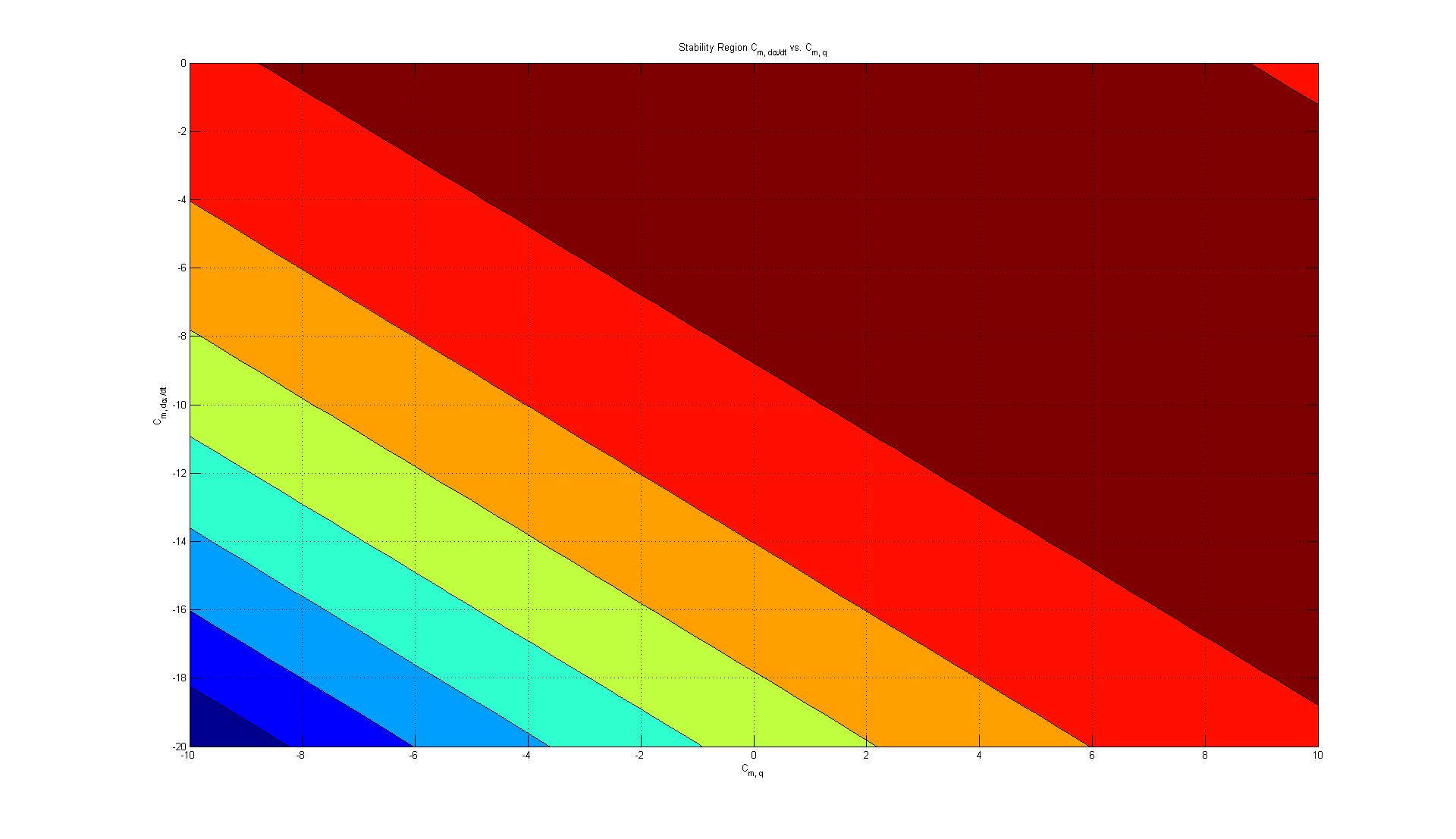}
\caption{\em Contour plot of stability region based on elevation $(C_{m_{\alpha}})$.}
\label{fig:contour1}
\end{center}
\end{figure}

\begin{figure}
\begin{center}
\includegraphics[scale=0.25]{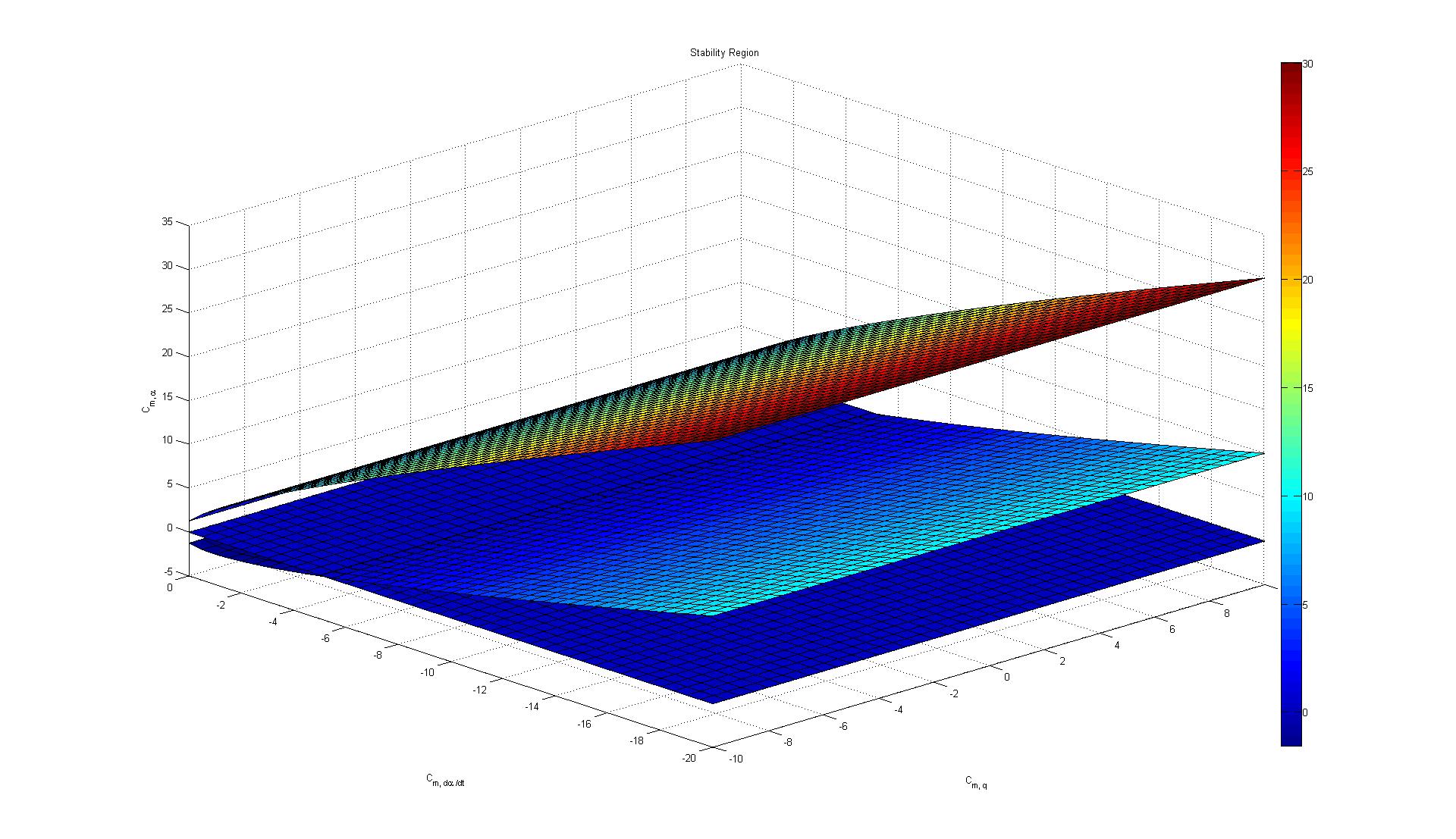}
\caption{\em Stability region bounded with upper and lower surface constraints.}
\label{fig:stabilitybounded}
\end{center}
\end{figure}

\begin{figure}
\begin{center}
\includegraphics[scale=0.3]{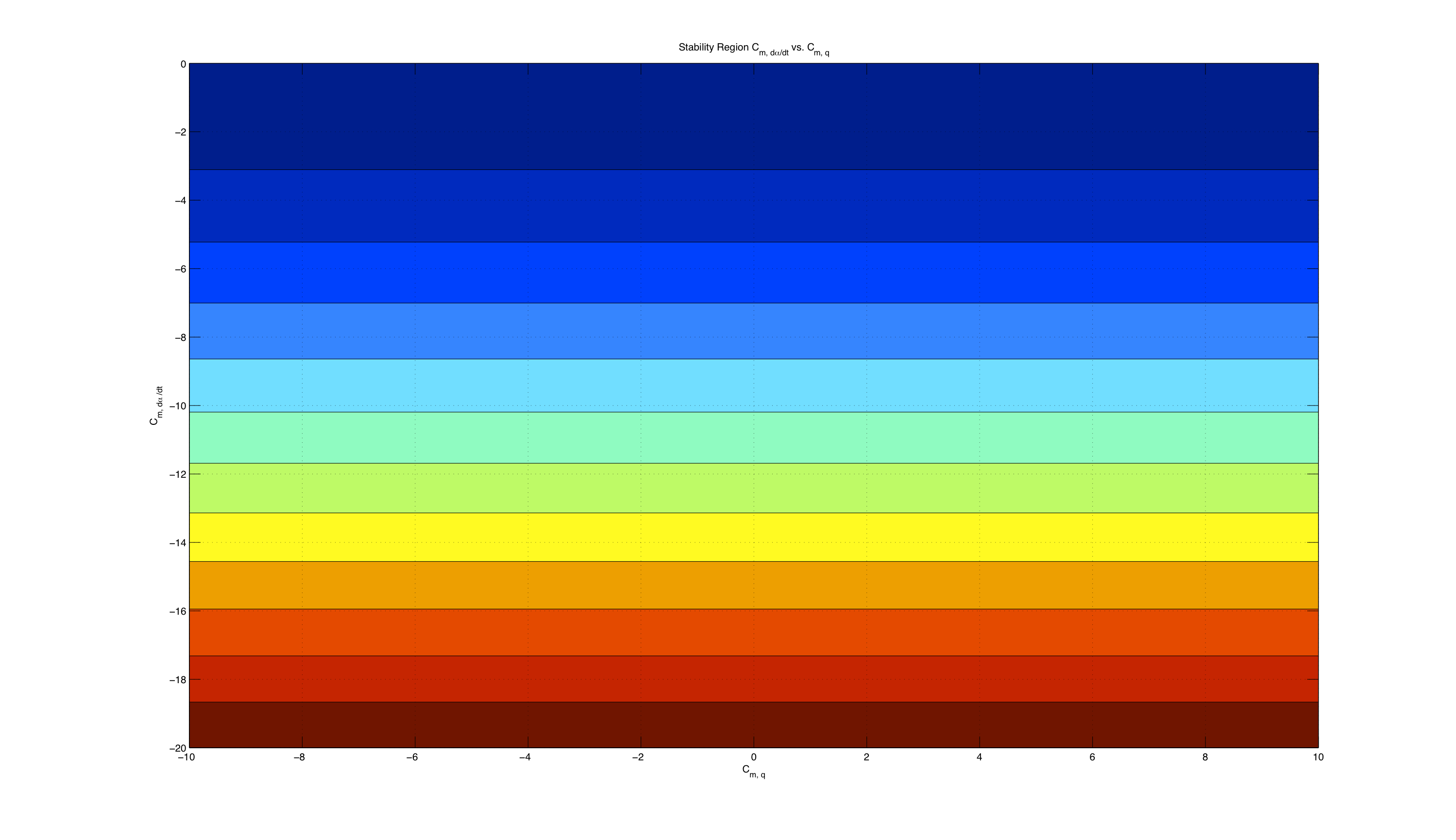}
\caption{\em Contour plot of stability region based on lower bound.}
\label{fig:contour2}
\end{center}
\end{figure}

\begin{figure}
\begin{center}
\includegraphics[scale=0.25]{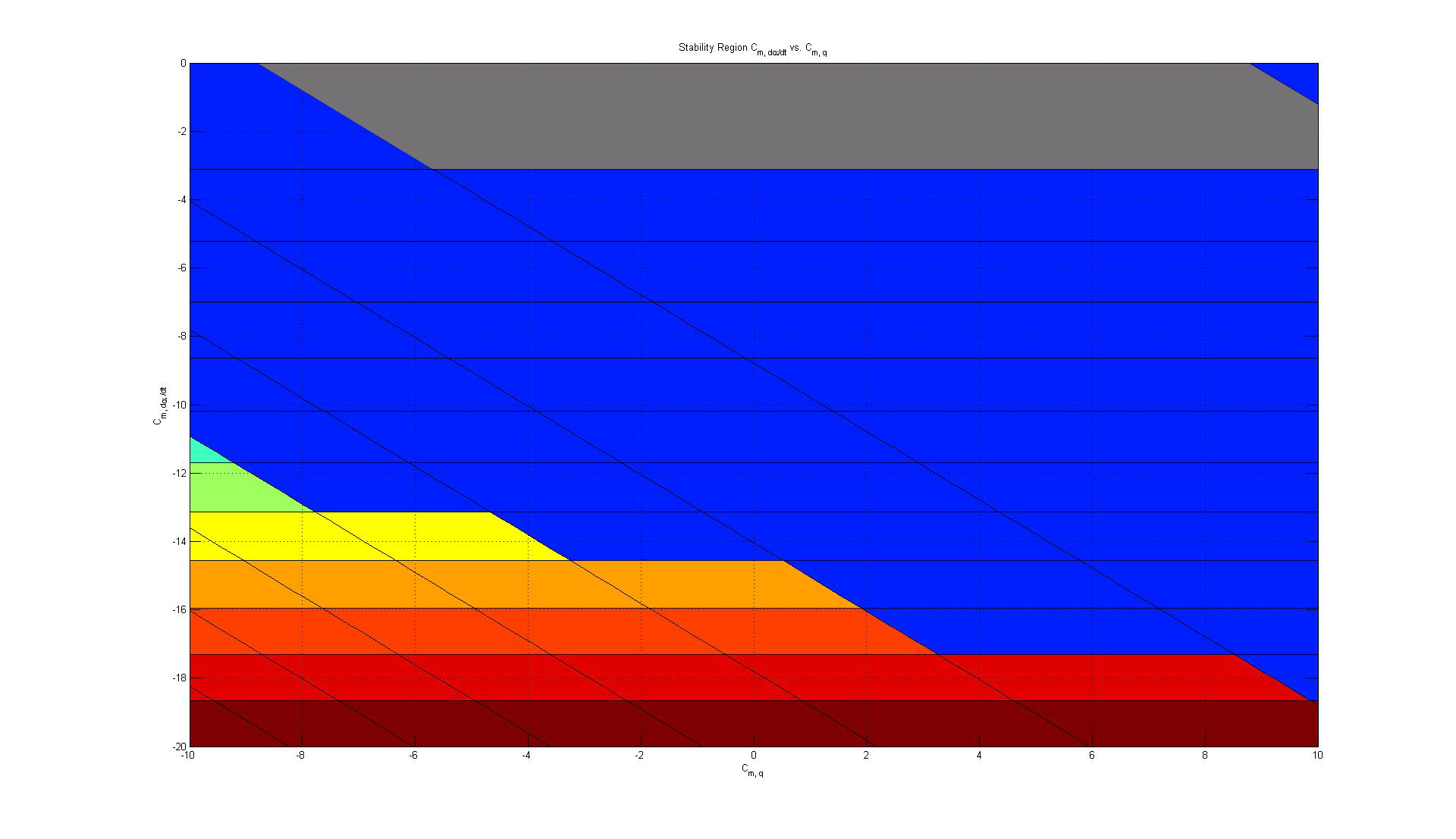}
\caption{\em Contour plot of stability region based on elevation $(C_{m_{\alpha}})$ with lower bound.}
\label{fig:contour3}
\end{center}
\end{figure}

The development of the stability region for the satellite in low elliptic Earth orbit followed the Lyapunov Lemma is stated. Based on this, Eq. \eqref{bbfinal} was integrated over $0$ to $\hat{T}$. Hence, 
\begin{align}\label{bint}
b(t)&=-2A\frac{C_{m_{\alpha}}}{l_{ref}}\int_0^{\hat{T}} \! {\rho(t)[v(t)]^2dt}+
\frac{A}{2}(C_{m_q}+C_{m_{\dot{\alpha}}})\int_0^{\hat{T}} \! {\frac{d}{dt}[v(t)\rho(t)]dt} \\
&-\frac{A^2}{4}(C_{m_q}+C_{m_{\dot{\alpha}}})^2\int_0^{\hat{T}} \! {[v(t)]^2[\rho(t)]^2dt} \nonumber
\end{align}
where $\int_0^{\hat{T}} \! {\rho(t)[v(t)]^2dt}=I_1$ and $\int_0^{\hat{T}} \! {[v(t)]^2[\rho(t)]^2dt}=I_2$  are constants. Eq. \eqref{bint} can be reduced to 
\begin{align}\label{bsimplified}
b(t)=-2A\frac{C_{m_{\alpha}}}{l_{ref}}I_1+\frac{A}{2}(C_{m_q}+C_{m_{\dot{\alpha}}})[v(\hat{T})\rho(\hat{T})-v(0)\rho(0)]-\frac{A^2}{4}(C_{m_q}+C_{m_{\dot{\alpha}}})^2I_2
\end{align}
For simplicity, let $k_1=-2\frac{A}{l_{ref}}I_1$, $k_2=\frac{A}{2}[v(\hat{T})\rho(\hat{T})-v(0)\rho(0)]$, and $k_3=-\frac{A^2}{4}I_2$ such that $k_1C_{m_{\alpha}}+k_2(C_{m_q}+C_{m_{\dot{\alpha}}})+k_3(C_{m_q}+C_{m_{\dot{\alpha}}})^2$ is the simplified expression for $b(t)$.

In order to successfully find the constants $I_1$ and $I_2$, some method of numerical integration is needed. Typically, integrations can be solved analytically if an expression exists and the integrand is integrable.  In this case, a numerical method was used for the elliptic sine approximations for the density and speed which posed a problem to evaluating a conventional integral. Necessary numerical approximation methods for these integrals were considered, such as Boole's rule and the trapezoidal rule.

Boole's rule is a variation on the Newton-Cotes' formula developed by George Boole \cite{Chapra}. This method of numerical integration approximates the integral of the type
\begin{equation}\label{Boole}
\int_{x_1}^{x_5}f(x)dx
\end{equation}
This method evaluates the integral by using values of $f$ over five equally spaced steps of $x_1$, $x_2=x_1+h$, $x_3=x_1+2h$, $x_4=x_1+3h$, and $x_5=x_1+4h$. Eq. \eqref{Boole} now exists as
\begin{equation}
\int_{x_1}^{x_5}f(x)dx=\frac{2h}{45}\Big[7f(x_1)+32f(x_2)+12f(x_3)+32f(x_4)+7f(x_5)\Big]+E
\end{equation}
where the error term is $E=-\frac{2^3}{1*3*5*7*9}h^7f^{(6)}(c)$ for some number $c\in[x_1,x_5]$.

The trapezoidal rule is a method of numerical integration where a finite number of trapezoids are constructed beneath the curve of the function of interest. The method involves finding the area of each of these trapezoids and then summing up these areas to find an approximation of the total area under the curve.\

This is accomplished by using Eq. \eqref{trap}, which is a variation on the simple equation for the area of a trapezoid. The error is minimized in this case, by maximizing the number of points that are used to do the approximation. In this case, the researchers used 502 points, matching the number of data points from the original density and altitude data.
\begin{equation}\label{trap}
\int_a^bf(x)dx\approx(b-a)\frac{f(a)+f(b)}{2}
\end{equation}
Like all numerical integration techniques, an error exists. An estimation for the local truncation error of a single application of the trapezoidal rule is $-\frac{1}{12}f''(\xi)(b-a)^3$ where $\xi$ lies somewhere in the interval from $a$ and $b$. This error term indicates higher order functions (with curvature) might consist of some error, while a linear function yields no error. The calculation of $I_1$ and $I_2$ utilized the trapezoidal rule for its simplicity.

The first constraint, $C_{m_{q}}+C_{m_{\dot{\alpha}}}=-10$, is a first order guess such that the pitch damping is similar. Pitch damping of flared projectiles has been investigated by Weinacht and co-authors \cite{PPT}. The study assumed
\begin{itemize}
\item long and slender axi-symmetric shape with a sharp nose
\item flight at sea level at speeds of 680 and 1700 $m/s$ (Mach 2 and 5)
\item zero-spin coning motion, i.e., coning motion with null spin rate
\end{itemize}

The Dipping Thermospheric Explorer (DipTE), however, is short and squat with a blunt nose and flies in a elliptic orbit at speeds of 7366 and 7950 $m/s$. With an aspect ratio $(L/D)$ of 3, and ratio of the center of gravity to length $(x/L)$ of 0.5, the ratio $x/D$ is 1.5. From Fig. ~\ref{fig:damping} it is determined that $C_{m_{\dot{\alpha}}}\approx0$ and $C_{m_{q}}\approx-10$. This is under the assumption that the center of mass is at the geometric center, i.e. the worst case scenario. This constraint is applied to Eq. \eqref{bsimplified} to determine an upper surface boundary under the Lyapunov Lemma. If the constraint is applied to the simplified form of $b(t)$, then $k_1C_{m_{\alpha}}-10k_2+100k_3 \leq \frac{4}{\hat{T}}$ if $k_1>0$. Therefore, it is determined that $C_{m_{\alpha}} \geq -0.0283792$. The representation of Fig.~\ref{fig:condition1} displays the stability region of applying the first constraint. It can be seen that the blue plane represents a graphical model of this constraint, as the magenta plane represents the boundary plane as discussed. Therefore, any values above the magenta plane would yield stability as the values below are unstable.

The second constraint assumes that $C_{m_{q}}$ and $C_{m_{\dot{\alpha}}}$ are free variables because the stability of the satellite depends on these coefficients. In this case, if their summation is equivalent to an arbitrary variable $w$, then $f(w)=k_1C_{m_{\alpha}}+k_2w+k_3w^2 \leq \frac{4}{\hat{T}}$ assuming $k_3 > 0$. The necessary condition to have is $\Delta={k_2}^2-4k_3(k_1C_{m_{\alpha}}-\frac{4}{\hat{T}})>0$ which may be different from the first condition. This condition exists due to the quadratic nature of the inequality and then proves and provides the lower boundary such that $C_{m_{\alpha}} \leq -0.0267126$. Therefore, the stability region will need to fall below this value as well.

Fig.~\ref{fig:contour1} demonstrates a graphical representation of the comparison of $b(t)$ to the criterion $\leq \frac{4}{\hat{T}}$ using MATLAB to ensure the bounded nature of Hill's equation. This figure shows a two-dimensional representation of the stability region of the satellite in terms on the pitch rate and pivot coefficients. This graphic is essentially a contour plot of the figure based on elevation from the three-dimensional model as in Fig.~\ref{fig:stability}. The multiple colors are generated by MATLAB to show this difference in graphical altitude, where the lighter colors (dark blue, blue, light blue, yellow-green, orange etc.) represent the lower regions and the darker red colors represent the higher regions. Specifically in this case, we are permitted immediately to examine the stability region of the satellite and associate it with the dark red region. This region falls between the upper and lower bounded surfaces of the criterion associated with the transformation of Hill's equation.

Fig.~\ref{fig:stabilitybounded} shows a three-dimensional representation of the region in which the satellite is expected to be stable and by default the remaining region where the satellite is expected to be unstable. The figure is made up of three distinct surfaces. The curved surface, although appears flat on this scale, corresponds to the full domain and range of the stability coefficients $C_{m_{q}}$, $C_{m_{\alpha}}$, and $C_{m_{\dot{\alpha}}}$. This is the full range of possibilities for the combination of these coefficients as the satellite traces its orbit and as limited by the physical limitations of the satellite as mentioned in the previous section. This figure shows an upper and lower curved boundary which displays the stability region lying between the surfaces. These boundaries were developed from the second constraint based on the quadratic expression $f(w)$ used with the criteria based on the Lyapunov Lemma. As previously discussed, a $C_{m_{\alpha, max}}$ and $C_{m_{\alpha, min}}$ (-0.0267126 and -0.0283792 respectively) were determined. This range was used in order to develop the roots of the quadratic expression such that 

\begin{equation}\label{roots}
w_{1, 2}=\frac{-k_2\pm\sqrt{\Delta}}{2k_3}
\end{equation}
where $\Delta$ is a function of $C_{m_{\alpha}}$, correspondingly making the roots $w_{1, 2}$ functions of $C_{m_q}$. Since the roots do exists, i.e. $f(w)<0$ if and only if $w_1<w<w_2$. Likewise, $C_{m_{q, top}}<w_2-y$ and $C_{m_{q, bot}}>w_1-y$. Therefore the surface as seen in Fig.~\ref{fig:stability} has a small portion representing the range and domain of the stability coefficients that would allow the satellite to remain stable between the boundary surfaces $C_{m_{q, top}}$ and $C_{m_{q, bot}}$.

Since it is evident that the curved surface provided in Fig.~\ref{fig:stability} intersects the lower surface, a contour plot can be developed to provide the region of stability. In Fig.~\ref{fig:contour2}, this plot gives elevation data for the lower boundary surface. The portion that the stability curve can be accounted for is the first dark blue region. The remaining colors are unstable regions due to falling below the lower bound. When combining the contour plots developed in Fig.~\ref{fig:contour1} and Fig.~\ref{fig:contour2}, a final representation of the stability region based on elevation with the lower bound included can be shown by Fig.~\ref{fig:contour3}. The gray region shows where all the conditions are met for stability of the satellite.

\section{Conclusion}
In this paper we used analytical and numerical methods and determined the stability
regions of the equation describing the one degree of freedom attitude dynamics 
in low altitude elliptic orbits.
The time dependent coefficients of the second order non-homogeneous ODE which describes the motion had a double periodic shape.
Hence, to approximate them we used a novel and powerful technique based on Jacobi elliptic functions using Jacobi elliptic sine function. Through a change
of variable the original ODE  which described the motion of the satellite was be converted into Hill's ODE suitable for stability analysis using Floquet theory. This allowed us to establish how changes in the coefficients of the ODE affect the stability of the solution via all the transformations. The expected result was be an allowable range of parameters for which the motion is dynamically stable or unstable. A possible extension of the application is a computational tool for the rapid evaluation of the stability of entry or re-entry vehicles in the rarefied flow regimes.

\end{document}